%% file: PranavetalV6.tex
\pdfoutput=1
\documentclass[10pt]{article}
\input{newcommands}
\usepackage{subfig}
\usepackage{amsmath,physics}
\usepackage{lineno}
\usepackage{enumerate}
\usepackage{enumitem}
\usepackage{tikz}
\usepackage{float}
\usepackage{graphicx}
\usepackage{natbib}
\usepackage{amsfonts,bm}
\usepackage{amssymb}
\usepackage{latexsym}
\usepackage{fontenc,comment}
\usepackage{rotating,booktabs}
\usepackage{verbatim}
\usepackage{array}
\usepackage{epstopdf}
\usepackage{cleveref}
\usepackage{setspace}
\usepackage{mathptmx}
\usepackage{newtxtext}
\usepackage{newtxmath}

\usepackage{ifthen}
\newboolean{dc}   
\newcommand{\volumef}{{\ooalign{\hfil$V$\hfil\cr\kern0.08em--\hfil\cr}}}

\setboolean{dc}{false}               
\graphicspath{{./}{./figures/domain/}{./figures/p_body/}{./v_compare/}{./figures/drag_plots/}{./figures/omgx_time/}{./figures/Hof_plots/}{./figures/symmetric_shedding/}{./pressure_coef/}{./figures/omgx_planes/}}   

\definecolor{sepia}{RGB}{180, 100, 20}
\definecolor{yellowgreen}{RGB}{113,124,18}
\definecolor{darkblue}{RGB}{40, 0, 128}
\definecolor{darkgreen}{RGB}{13, 102, 13}
\definecolor{edred}{RGB}{255, 0, 50}

\newcommand\Pranav[1] {{\color{darkgreen}{#1}}}

\newcommand\Geno[1] {{\color{black}{#1}}}
\newcommand\Sutanu[1] {{\color{black}{#1}}}
\newcommand\C[1] {{\color{black}{#1}}}
\newcommand{\myabstract}  
{
}

\begin{document}

%
%

%
%

\title{\textbf{\large{ 
High drag states in tidally distorted wakes
}}}
%
%
\author{\textsc{Pranav Puthan }\\
\textit{\footnotesize{Department of Mechanical and Aerospace Engineering, University of California, San Diego}}\\
				\textsc{Geno Pawlak, Sutanu Sarkar}
				\thanks{\textit{Corresponding author address:} 
				Department of Mechanical and Aerospace Engineering, University of California, San Diego, 
				9500 Gilman Dr., La Jolla, CA 92092.}\\
\textit{\footnotesize{Department of Mechanical and Aerospace Engineering, University of California, San Diego}}\\
\textit{\footnotesize{ and Scripps Institute of Oceanography, La Jolla}}
}

\maketitle 

\ifthenelse{\boolean{dc}}
{
\twocolumn[
\begin{@twocolumnfalse}
\amstitle

\begin{center}
\begin{minipage}{13.0cm}
\begin{abstract}
	\myabstract
         \newline
	\begin{center}
		\rule{38mm}{0.2mm}
	\end{center}
\end{abstract}
\end{minipage}
\end{center}
\end{@twocolumnfalse}
]
}

\onecolumn 

\abstract{
 Large eddy simulations (LES) are employed to investigate the role of time-varying currents on  the form drag and vortex dynamics of \Sutanu{submerged}  3D topography in a stratified rotating environment. 
 The  current is of the form  $U_c+U_t \sin(2\pi f_t t)$, where $U_c$ is the mean, $U_t$ is the tidal component and $f_t$ is its  frequency.  A conical obstacle is considered in the regime of low Froude number. When tides are absent, eddies are shed at the natural shedding frequency $f_{s,c}$. 
The relative frequency  $f^*=f_{s,c}/f_t$ is varied in a parametric study which reveals
states of high time-averaged form drag coefficient.
There is a two-fold amplification of the form drag coefficient  relative to the no-tide ($U_t=0$) case when $f^*$ lies between 0.5 and 1. 
  The spatial organization of the near-wake vortices in the high drag states is different from a \Karman{} vortex street.
For instance, the vortex shedding from the obstacle is symmetric when $f^*=5/12$ and
strongly asymmetric when $f^*=5/6$. The increase in form drag \Sutanu{with increasing $f^*$} stems from bottom intensification of the pressure in the obstacle lee which is  linked to changes in  flow separation and near-wake vortices. 
}

\section{Introduction} \label{intro}

Rough bottom topography in the abyssal ocean contributes significantly to enhancement of drag and turbulent dissipation.
\cite{EgbertR_N:2000,EgbertR_N:2001} estimate that up to 1 TW of power is lost \textit{in-situ} from tide-topography interactions in the abyssal ocean. 
When abyssal flow encounters rough topography, energy is lost 
in two ways: (1) through skin friction resulting from tangential stress at the boundary and (2) via pressure/form drag resulting from normal stress. Energy loss from friction is usually small, with estimates in the $\mathcal{O}(0.02$ mW/m$^{2}$) or approximately 7 GW on a global scale \citep{jayne01}, accounting for less than 1\% of the 1 TW estimate of \cite{EgbertR_N:2000,EgbertR_N:2001}. {Recent studies of \cite{ZhangM_JGR:2020} and \cite{Klymak_JPO:2018} highlight the crucial role played by \Sutanu{multiscale} topography in extracting momentum (through topographic form stress) from the background flow and maintaining a dynamic balance in the abyssal ocean}. Thus, form drag is often the primary mechanism of energy extraction from the barotropic tide, especially at steeper topographies \citep{McCabePP_JPO:2006,HowritzTL_CSR:2021}. 

   In situ measurements in literature show  that the loss of momentum associated with form drag 
    is enhanced by obstacles in the coastal ocean such as headlands \citep{EdwardsMM_JPO:2004,Magaldi_OM:2008,WarnerMMN_JPO:2012,WarnerM_JGR:2014} 
   and continental shelves \citep{NashM_JGR:2001,Wijesekera_JPO:2014}.
   However, observational studies of form drag estimates from the abyssal ocean are limited. Lack of information on \C{the} magnitude and spatial distribution of form drag presents a challenge for form drag parameterizations in global climate models (GCMs). Numerical studies can play an important role in bridging this gap. \cite{WarnerM_JPO:2009} performed numerical simulations \C{with the hydrostatic ROMS model} of a non-rotating tidal flow past a Gaussian headland to examine different components of form drag. 
   \Sutanu{
   They showed that while the normalized separation drag  (the average drag coefficient) increased with an increase in the aspect ratio of the headland, it does not depend on the tidal excursion or the headland size.
   }
  In the present work, we examine form drag in tidally modulated flow past an underwater obstacle. Turbulence resolving simulations enable us to study the time varying flow past obstacles without compromising on the accuracy of their representation. The characterization of flow separation and pressure distribution on the obstacle allows us to link underlying physical mechanisms to any changes in the observed form drag. 
  %

   Form drag on an obstacle is dependent on the ambient stratification. 
   \C{When} a steady current encounters a 3D ridge, the flow transitions to a state of high drag when the Froude number reduces below 1, \C{e.g. \cite{EpifanioD_JAS:2001,VosperCSM_JFM:1999}. }
%
%
%
In situ tidal measurements of form drag are challenging and there are few such observations, e.g. \cite{VoetAMN_JPO:2020}, \C{who infer pressure from density measurements using the hydrostatic approximation.}
Additionally, tide induced unsteadiness may lead to changes in flow separation and distribution of lee vorticity. For example, tidal currents create transient lee eddies (or lee vortices) in wakes behind headlands \citep{PawlakME:2003,CallendarKF_OS:2011,MacKinnonA_JGR:2019} and submerged topography \citep{GirtonMZA:2019}.
 Complexity in the impinging flow, variable stratification  and irregular bathymetry at these sites present a challenge in elucidating the role of lee eddies. 
To examine the role of tides in flow separation and form drag, we perform turbulence-resolving simulations of an oceanic wake past a conical hill generated by a tidally modulated flow. The background flow may be expressed as $U_b=U_c +U_t \sin(\Omega_t  t)$, where $U_c$ and $U_t$ are the mean and tidal components and $\Omega_t=2\pi f_t$  is the tidal frequency (in rad/s). 

Stratification, rotation and tidal forcing 
are key elements of geophysical wakes. When a tidally modulated flow encounters an underwater obstacle, the wake structure is governed predominantly by 
 the obstacle Froude number ($Fr_c$), the tidal excursion number ($Ex_t$), the Rossby number ($Ro_c$) and velocity ratio ($R$) :
 \begin{align*}
 Fr_c=\frac{U_c}{Nh} \qquad ; \qquad Ex_t=\frac{U_t}{\Omega_t D} \qquad ; \qquad Ro_c=\frac{U_c}{f D} \qquad ; \qquad R=\frac{U_t}{U_c} \; , 
\end{align*}  
where $N$ is the background buoyancy frequency, $f$ is the inertial frequency, $h$ is the height of the obstacle and $D$ is the obstacle base diameter.
 In unstratified environments, a hill with a 3D geometry does not shed 
 vortices similar to the vertically coherent
  lee eddies which are observed in the ocean. 
 Instead a standing horseshoe vortex and periodic hairpin vortices are observed downstream at low Reynolds number ($Re_D$) 
\citep[see][]{AcarlarS_part1_1987} which become indistinct at higher Reynolds number \citep{hill_GarciaLRL_JFM:2009}. A low $Fr_c$ ($Fr_c \ll 1$) flow is constrained to move around rather than over the obstacle, owing to the large potential energy barrier. 
This leads to lateral shear layer roll up into lee vortices \citep{hill_HuntS_JFM:1980}.

Topographic wakes are also affected by planetary rotation. A \C{relatively} large planetary rotation rate \Geno{(small $Ro_c$)} 
 induces asymmetry in the strength of cyclonic and anticyclonic eddies shed from the topography \citep{DietrichBLM_GASF:1996}. \cite{DongMS_JPO:2006} attributed the loss in symmetry to centrifugal instabilities in the wake. \cite{hill_PerfectKR_GRL:2018} and \cite{hill_SrinivasanM_JPO:2018} showed that the change in vertical structure of wake vortices is governed by the Burger number $Bu$, defined as $(Ro_c/Fr_c)^2$. Their idealized simulations show decoupling of vortices along their vertical extent owing to loss of geostrophic balance, when $Bu>12$. 

The regime of weak rotation and strong stratification (or equivalently, large $Bu$ ) 
 applies to 
wakes behind  abyssal hills. For example, consider the abyssal hills in the Brazil Basin \citep{LedwellMPLS_N:2000,NikurashinL:2011}. 
The bottom topographic roughness is $\mathcal{O}(1$ km) in the horizontal. For an obstacle with $D=1.5$ km, buoyancy period of 1 hr and $U_c=U_t=10$ cm/s, the tidal excursion number is $Ex_t\approx 0.5$ for \C{the} M2 tide and the average value of $Fr_c$ lies close to 0.2. The value of Rossby number is $Ro_c\approx$ 3, at 15$^\circ$S latitude. The resolution of GCMs is insufficient to resolve these hills. Thus, parametrization of the wake dynamics at these length scales is critical.   

Owing to numerical constraints, idealized simulations often ignore tidal forcing. Yet, in situ observations affirm that tides can significantly influence  flow separation at islands, continental slopes and submerged topography.
Observations by \cite{BlackG_JGR:1987} showed the formation of `phase' eddies in the continental shelf of Great Barrier Reef. 
 \cite{DennissMM_JGR:1995} and \cite{ChangJLCM_JPO:2019} reported lee eddies shed past islands at the dominant tidal frequency. This phase-locking phenomenon is observed even when the tidal velocity amplitude is small relative to the mean flow. 
Recently \cite{PuthanGRL:2020} \C{found tidal synchronization in a study of flow past a conical hill where the frequency of the far-wake lee vortices locked to a subharmonic (depending on the value of $f^*= f_{sc}/f_t$) of the tidal frequency. }
%
The relative frequency is linearly related to $Ex_t$ as $f^*=2\pi St_c Ex_t/R$, where $St_c$ is the vortex shedding Strouhal number in a steady background flow. This relation simplifies to $f^*=1.66 Ex_t$ for the hill wake ($St_c=0.265$) \C{when} $R=1$. 
However, the near wake characteristics such as flow separation at the hill and the attendant form stress \C{were not studied by \cite{PuthanGRL:2020} and} have not received adequate attention in \C{other} previous studies.

%

In this work, 
\C{we} address the questions pertaining to momentum loss of abyssal currents during flow-topography interactions at the obstacle and the associated 
wake-vorticity distribution. 
 The motivation for this work is two-fold. \Geno{We} explore possible states of large form drag owing to changes in 
 pressure distribution in the lee 
 \Geno{and} determine the qualitative changes to the vorticity distribution in the near wake in each state. 
 \C{The} numerical formulation is detailed in \cref{subsec:comp_model}. \Cref{subsec:param} introduces the parameter space and lists the cases performed in the study. A brief introduction to form stress and an overview of previous literature \C{related to form drag} is provided in \cref{sec:fdrag_theory}.
 \Cref{subsec:pres_anomalies} elucidates the changes in form drag on varying $f^*$ and \cref{subsec:vortex_dynamics} elucidates the underlying changes to flow separation. 
 The paper concludes with a brief summary in \cref{sec:conclude}.

\begin{figure}[ht!]
\centering
\includegraphics[trim={0cm 0cm 0cm 0cm},clip, width=0.9\linewidth, angle=0]{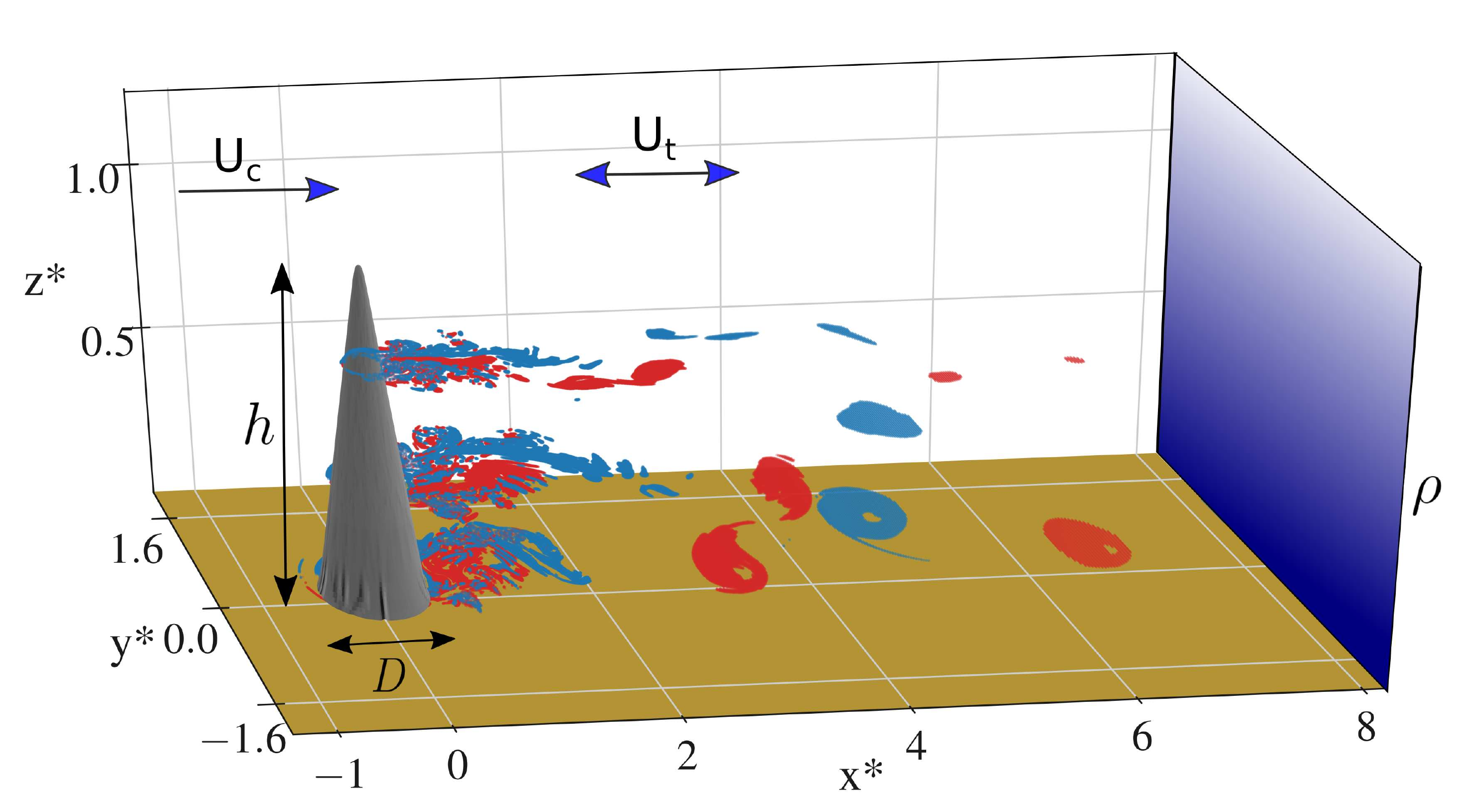}
\caption{
\C{The numerical model setup is shown for the following model problem:} a tidally modulated current encounters a conical obstacle in a stratified environment. The vortex shedding pattern in three horizontal planes for case $f^*=2/15$ is represented using isosurfaces of vertical vorticity corresponding to $\omega_z/f=\pm 10$.
}
\label{fig:domain}
\end{figure}

 \section{Computational model}\label{subsec:comp_model}
\input{Computational_model}


\begin{table}[]
\centering
\small
\begin{tabular}{@{}|c|c|c|c|c|@{}}
\toprule
\textbf{Regime} & \textbf{Case} & \textbf{Category}         & \textbf{\begin{tabular}[c]{@{}c@{}}Far wake\\ shedding \\ frequency($f_s$)\end{tabular}} & \textbf{\begin{tabular}[c]{@{}c@{}}Near wake\\ vortex shedding \\ pattern \end{tabular}}      \\ \midrule
1      &  $f^*=\infty$       & No tides           & $f_{s,c}$                                                                                & K\'arm\'an vortex street                 \\ \midrule
2      & $f^*=2/15$         & $f^*<1/4$             & $f_{s,c}$                                                                                & Vortex pulses + K\'arm\'an vortex street \\ \midrule
3     & $f^*=5/12$           & $1/4 \leq f^*< 1/2$    & $f_t/4$                                                                                  & Symmetric twin dipoles                 \\ \midrule
4    & $f^*=5/6$            & $1/2 \leq f^* \leq  1$ & $f_t/2$                                                                                  & Strong asymmetric shedding             \\ \bottomrule
\end{tabular}
\caption{Different cases in this study: The relative frequency $f^*=f_{s,c}/f_t$ is varied among 9 cases.
 \C{Four regimes with different patterns of wake vortices are observed and are discussed with representative cases shown in the table.}
  The values of  $Re_D$, $Fr_c$ and $Ro_c$ are fixed at 20000, 0.15 and 5.5 respectively.}
\label{tab:cases}
\end{table}

\section{Form drag in oscillatory flows} \label{sec:fdrag_theory}
Flow-bathymetry interactions result in pressure differences across abyssal obstacles, which manifests as form drag. 
 %
 Recent measurements at Palau estimated large form drag 
 acting on the tidal and mean components of the flow at very low Froude numbers 
\citep{VoetAMN_JPO:2020}. 
 The drag force due to form drag is computed as 
\begin{equation}
F_D = -\int p \delta_{1j} n_j dA
\label{eq:drag}
\end{equation}
Using \cref{eq:pres_decompose},  \cref{eq:drag} may be \Geno{expanded} as 
\begin{align}
F_D = -\int p_d \delta_{1j} n_j dA + \rho_0 \volumef \dv{ U_b }{t}
\label{eq:drag3}
\end{align}
The first term on the right-hand side of \cref{eq:drag3} includes contribution from two sources : (a) the separation drag associated with flow separation at the obstacle ($F_D^S$) and (b) the added mass ($F_D^A$) \citep{lamb:1930,KeuleganC:1958}. 
The drag associated with added mass is approximated in literature  \citep[e.g.][]{Morison:1950} using the relation
\begin{equation}
F_D^A = C_a\rho_0 \dv{U_b}{t} \volumef
\label{eq:added_mass}
\end{equation}
where, $C_a$ is the added mass coefficient. The second term on the right-hand side of \cref{eq:drag3} represents the Froude-Krylov force ($F_D^{FK}$) associated with the oscillating pressure gradient $p_\infty$ \citep{YuRH_JGR:2018}. 
The sum of the added mass force ($F_D^A$) and the Froude-Krylov force ($F_D^{FK}$) is referred to as inertial drag ($F_D^I$).
\begin{align}
 F_D^I &= F_D^{FK} + F_D^A = \rho_0 \dv{ U_b }{t} \volumef + C_a  \rho_0 \dv{ U_b }{t} \volumef   \label{eq:drag_inertia0}   
 \end{align}
 Therefore, the total drag force in \cref{eq:drag3} can now be written \C{as} 
  \citep[e.g.][]{WarnerM_JPO:2009} :
 \begin{align}
 F_D &= F_D^S + F_D^A + F_D^{FK} =  F_D^S + F_D^I
\end{align}

 \cite{WarnerM_JPO:2009} concluded that \C{the} net contribution of inertial drag 
vanishes when averaged over a tidal cycle. 
Therefore, its contribution is estimated using \cref{eq:drag_inertia0}, and removed from $F_D$ to compute the separation drag.
The separation drag is assumed to be in-phase with $U_b$ while the inertial drag variability \C{has} a phase difference of $\pi/2$ with respect to $U_b$ \citep{Morison:1950}.
 
 A similar drag force decomposition was proposed by \cite{Lighthill_JFM:1986}, wherein 
 the inviscid inertial drag is estimated from potential flow theory. 
However, the force decomposition of \cite{Lighthill_JFM:1986} has limitations. \cite{Sarpkaya_JFS:2001} argued that the force decomposition of \cite{Lighthill_JFM:1986} might exclude the effects of
 viscosity on the added mass. 
 His results also demonstrate that the added mass coefficient $C_a$ is time-dependent.  
  He asserted that 
the  subtraction of the ideal inertial force (calculated from potential flow approximation) from the total form drag, leaves behind a force that consists of both the separation drag \textit{and} an `acceleration-dependent' inertial force.
%
%
However, the contribution from inertial drag is negligible in the  time-averaged value of  $F_D$, when the average is computed over a large time duration (spanning multiple tidal cycles) in a statistically stationary flow  \citep{Sarpkaya_JFS:2001}. Thus the approximations of \cite{WarnerM_JPO:2009} can serve as a theoretical basis to aid the interpretation of the present results by providing \C{an adequate}  estimate of the separation drag. In other words, the time-mean of \Geno{the} form drag is representative of 
the separation drag $F_D^S$. 

In the next section, we examine 
the time-averaged mean, phase averaged and root-mean-squared (RMS) values of $F_D$ defined below:
\begin{align*}
  &\text{Mean drag: } \langle F_{D} \rangle =  \frac{ \int_0^{nT} F_D dt }{nT} \\
  &\text{Phase averaged drag: } \langle F_{D}\rangle_\phi  =  \dfrac{ \sum\limits_{k=0}^{n}  F_D(\Omega_t T k + \phi) }{n} \\
  &\text{RMS drag: } \langle F_{D} \rangle_{rms} =  \sqrt{ \frac{ \int_0^{nT} (F_D - \langle F_D \rangle)^2 dt }{nT} }
\end{align*}
  Here, $T$ is the tidal period, \Geno{$\phi=\Omega_t t$ is the tidal phase} and $n$ is the number of tidal cycles. The value of $n$ is larger than 6 for all cases.
%



\section{Pressure anomalies in the wake } \label{subsec:pres_anomalies}
\input{Pres_anomalies}

\section{Vortex dynamics} \label{subsec:vortex_dynamics}
\input{vortex_dynamics}




\subsection{Symmetric eddy dipoles} \label{subsubsec:symm_shedding} 
\input{symmetric_vortices}


\subsection{Temporal variation of vorticity} \label{subsubsec:temporal_vort}
\input{Temporal_vort}





\section{Discussion and conclusions} \label{sec:conclude}
LES were undertaken to examine the wake created by a stratified tidally modulated flow ($U_b$) in the presence of weak background rotation. The barotropic flow $U_b$ has a mean component $U_c$, and a tidal component $U_t$ of equal strength. 
\C{Since} the Froude number $Fr_c$ is small, the near bottom flow is forced to separate laterally from the obstacle. As a result, coherent eddies of the nature observed in the ocean \citep[e.g][]{PawlakME:2003,MacKinnonA_JGR:2019}, form in the lee. 
%
We find that the flow exhibits four regimes, based on vortex patterns, as summarized in \cref{tab:cases}. 
In the first regime where tides are absent, lee vortices separate from the obstacle at a constant frequency $f_{s,c}$ and form a \Karman{} vortex street downstream. In the next three regimes, the arrangement of vortices is altered by the tidal flow. 
Changes in flow separation, accompanied by bottom-intensified pressure differences on the obstacle, are responsible for states of high drag in  \C{the tidally modulated cases, especially} regime 4.

\C{The effect of tidal oscillations is characterized}
  by varying the relative frequency parameter ($f^* = f_{s,c}/f_t$) from $1/10$ to 1. 
 Three of the nine cases are chosen, namely, $f^*=2/15, 5/12$ and $5/6$ from regimes 2, 3 and 4, respectively, to illustrate \C{the results}. 
 At $f^*=2/15$ (regime 2), vortex pulses which occur every tidal cycle, feed vorticity into a larger eddy in the recirculation zone. These larger eddies form a \Karman{} vortex wake downstream.  At $f^*=5/12$ (regime 3), laterally symmetric eddy dipoles are shed in the near wake. The lateral symmetry is controlled by the tidal flow up to a streamwise distance of $U_t/f_t$. Beyond this location, the vortices partially break down or merge to create a disorganized wake downstream. This event is accompanied by a change in the wake vortex frequency from $f_t$ to $f_t/4$. \C{At} $f^*=5/6$ (regime 4), strong asymmetric shedding at a frequency of $f_t/2$ is observed in the entire wake. The wake is laterally wider in comparison to the other regimes.
%

 The timing of the shed vortices is strongly influenced by the barotropic tidal oscillation. 
  When  $|\omega_z|$ is volume-averaged over the wake, its \C{temporal} variation is affected by the tidal oscillation. At $f^*=2/15$, the temporal evolution is in phase with tides owing to the formation of vortex pulses. On the other hand, at $f^*=5/12$, the temporal evolution is out of phase with the tidal oscillation. Excess vorticity is added to the wake from the lateral motion of recirculating fluid and the attendant shear-layer roll-up during the low-velocity phase.

 Changes in flow separation also lead to variations in pressure along the streamwise direction of the obstacle. The \C{difference between} the mean pressure field fore and aft of the hill ($\Delta \langle p_d \rangle$) is bottom intensified in cases $f^*=5/12$ and 5/6. 
 At $f^*=5/12$, the intensification is laterally offset from the centerline due to the impinging of the recirculating fluid on the obstacle centerline  near the zero-velocity phase. 
 
 The normalized form drag, i.e.  drag coefficient, obtained by integrating the pressure field over the obstacle surface area, varies amongst these cases. Form drag has two components, namely, the inertial and the separation drag. \C{The separation drag is the dissipative part of form drag.}
%
 The mean drag (\C{averaged over several cycles}) which is \Geno{associated with} the separation drag \citep{Sarpkaya_JFS:2004,WarnerM_JPO:2009}, generally increases 
 with increasing $f^*$. The bottom intensified values of 
 $\Delta \langle p_d \rangle$ (associated with large eddies) and the longer eddy residence time in the lee contribute to approximately a two-fold increase of mean drag \Sutanu{coefficient} in regime 4 relative to its value in the no-tide case. High drag states are also present in regime 3, wherein approximately a 60\% increase is observed in mean drag coefficient relative to the no-tide case. 
 Therefore, for a continuous distribution of topographic scales in the abyssal ocean, obstacles with $1/4 \leq f^* \leq 1$ \Geno{can} preferentially remove momentum from the background tidally modulated flow. The inertial drag force (associated with the magnitude of tidal acceleration) is dominant in regime 2, causing large variance of form drag within the tidal cycle in the $f^*=2/15$ case. Since the inertial component of drag decreases with increasing $f^*$, so does the RMS drag. 
 The rapidly varying pressure gradient associated with tidal forcing in $f^*=2/15$ restricts the size of the lee eddy and diminishes the pressure anomalies during the zero-velocity phase of the tidal cycle. This leads to lower mean form drag relative to regimes 3 and 4. 

 \Geno{\cite{WarnerM_JPO:2009} investigated a coastal headland in a purely tidal flow and did not find a significant effect of the excursion number (equivalently $f^*$) on separation drag. 
 Deviations \Sutanu{of the present findings (regarding excursion number effects)} from their results may be attributed to a difference in geometry (submerged conical obstacle) or the presence of a mean current in this study.}
 The drag coefficient associated with mean form drag normalized using the obstacle frontal area in \C{regime 4} is 3. Converting this to a value  based on the obstacle plan area, the drag coefficient takes a value of approximately 0.6, substantially larger than the drag coefficient associated with the frictional bottom boundary layer of $\mathcal{O}(10^{-3}$) estimated in previous work \citep[e.g.][]{DeweyC_JPO:1988}.    %
 %
 %
  %
 Thus, form drag {owing to wakes of steep obstacles dominates frictional drag in tidally modulated currents.}

\bibliography{mybibfile,Add_refs}

\end{document}

%% file: newcommands.tex
\newcommand\Karman{K\'arm\'an}

\newsavebox{\astrutbox}
\sbox{\astrutbox}{\rule[-5pt]{0pt}{20pt}}

\newcommand{\be}{\begin{equation}}
\newcommand{\ee}{\end{equation}}
\newcommand{\bee}{\begin{eqnarray}}
\newcommand{\eee}{\end{eqnarray}}

%% file: Computational_model.tex
The computational domain is 9.5 km in the streamwise ($x$) direction, 3.8 km in the spanwise ($y$) and 2 km in the vertical ($z$) direction. The conical obstacle of height $h$ and base diameter $D$ is placed at the origin. 
%
%
For convenience, the horizontal and vertical distances are normalised by $D$ and $h$ such that $x^*=x/D$, $ y^*=y/D $ and $z^*=z/h$. 
  The problem set-up is illustrated in \cref{fig:domain}. A barotropic unidirectional current $U_b=U_c + U_t \sin(2\pi f_t t)$ encounters a conical obstacle in a uniformly stratified environment.


The 3D non-hydrostatic flow is modeled by the Navier-Stokes equations under the Boussinesq approximation. The equations for conservation of mass, momentum and density on an \textit{f}-plane are given below in tensor notation: 
 \begin{align}
  \pdv{u_m}{x_m} &= 0  \label{eq:mass} \, , \\
  \pdv{u_m}{t} +  \pdv{(u_n u_m)}{x_n} -  f \epsilon_{mn3} (u_n - U_b \delta_{n1}) &= -\frac{1}{\rho_0} \pdv{p}{x_m} - \frac{g\rho'}{\rho_0} \delta_{m3} + \pdv{\tau_{mn}}{x_n}     \, , \label{eq:momentum} \\
   \pdv{\rho}{t} + \pdv{(u_n \rho)}{x_n} &= \pdv{\Lambda_n}{x_n} \, . \label{eq:density}
 \end{align}
where, $u_m=(u_1, u_2, u_3)=(u,v,w)$  denotes the velocity components and $\rho$ is the density field.
Here, $\rho'$ represent deviation of density from its background value and $p$ is the deviation from the mean pressure imposed by geostrophic and hydrostatic balance. 
The pressure deviation $p$ may be represented as:
\begin{equation}
p = p_d + (x-x_0) \pdv{p_\infty}{x}
\label{eq:pres_decompose}
\end{equation} 
where, $\partial p_\infty /\partial x = -\rho_0 dU_b/dt =  -\rho_0 U_t \Omega_t \cos(\Omega_t t)$ is the  pressure gradient driving the barotropic current $U_b$ and $p_d$ is the dynamic pressure.
%
 %
The stress tensor $\tau_{mn}$ and density flux vector $\Lambda_n$ are computed using the approach of \cite{PuthanOM:2020}. The grid parameters, time-advancement scheme and boundary conditions are adopted from \cite{PuthanGRL:2020}.
%

   \section{Simulation parameters} \label{subsec:param}
A regime of weak rotation and strong stratification is considered in the study. Inertial effects on the wake are weak in the lee of abyssal hills and headlands at length scales of $\mathcal{O}$(1km) near the equator \citep{Rudnick_O:2019,LiuC_JGR:2018}. These flows may be classified under the high Rossby number regime ($Ro_c >1$). 
  The inertial frequency is set to its value at $15^\circ N$ such that $Ro_c=5.5$. 
 \Geno{Much of the topography in the abyssal ocean is subject to flow conditions with $Fr_c \ll 1$
 e.g.  \cite{NikurashinF_JPO:2010}.}
   %
    To this end, we consider a topographic Froude number $Fr_c$ of 0.15 
where the flow is predominantly    
    around the obstacle, creating coherent vortices as observed in geophysical wakes \citep{hill_PerfectKR_GRL:2018}.

    %
       %
 Tidal modulations are added to the mean flow. \C{We consider tidal velocities of amplitude equal to 
 the mean current, 
 so that $R=U_t/U_c=1$.}
  The relative frequency $f^*=f_{s,c}/f_t$ is varied from 0.1 to 1 \C{in a parametric study with nine cases} to examine the variations in vortex dynamics and form drag. Equivalently, $Ex_t$ is varied from 0.06 to 0.6 (assuming $St_c=0.265$), values of relevance in the ocean \citep{SignellG_JGR:1991,EdwardsMM_JPO:2004,MustgraveMPW_JPO:2016}. At a constant tidal frequency $f_t$ e.g. the M2 tide, larger $f^*$ values are associated with obstacles of smaller length scales. 
  
   Different regimes of tidal synchronization were observed 
   by \cite{PuthanGRL:2020}, wherein the lee vortices in the far wake were found at frequencies $f_{s,c}$, $f_t/4$ or $f_t/2$. These regimes are listed in \cref{tab:cases}. 
   Regime 1 consists of a single case ($f^*=\infty$) with no tidal modulations in the background flow (i.e. $R=0$). In regime 2-4, $R$ equals 1 and multiple cases are explored \Geno{within} each regime at discrete $f^*$ values. As listed in the second column of \cref{tab:cases}, one case is chosen \C{to illustrate each regime: 
   $f^*=2/15,5/12 $ and 5/6 cases, for regimes 2, 3 and 4, respectively.}
   Each regime has a distinctive near-wake vortex shedding pattern (listed in the 5th column of \cref{tab:cases}). The differences in the vortex shedding at the hill play a crucial role in changing the topographic pressure distribution, motivating our study of 
   %
   %
   role of $f^*$ on \C{the} separation of vortices {at the obstacle} and 
   its influence on form drag. 
  
  

%% file: Pres_anomalies.tex

In a low $Fr_c$ environment, the encounter of a steady current with a 3D obstacle forces a significant volume of fluid to navigate laterally.
This laterally driven flow separates and  wake eddies are  
formed in the lee.
%
 The form stress increases as a result.
%
%
To examine the origin of form stress, a \C{thorough characterization} of the dynamic pressure $p_d$ is crucial. The form drag \C{is} computed \Geno{as the sum of} the surface integral of $p_d$ and the Froude-Krylov force, as shown in \cref{eq:drag3}. 
%
Note that the surface integral of $p_d$ is inclusive of the added-mass component of inertial drag. 
In \C{section 5a}, the dynamic pressure field is examined in the wake and on the obstacle. Changes in eddy shedding and form stress are illustrated and quantified over the flow regimes listed in \cref{tab:cases}.
Estimates of $F_D$ and their values relative to $f^*=\infty$ (no-tide) case, are provided in \C{section 5b}.  

\begin{figure}[ht!]
\centering
\includegraphics[trim={0cm 0cm 0cm 0cm},clip, width=0.85\linewidth, angle=0]{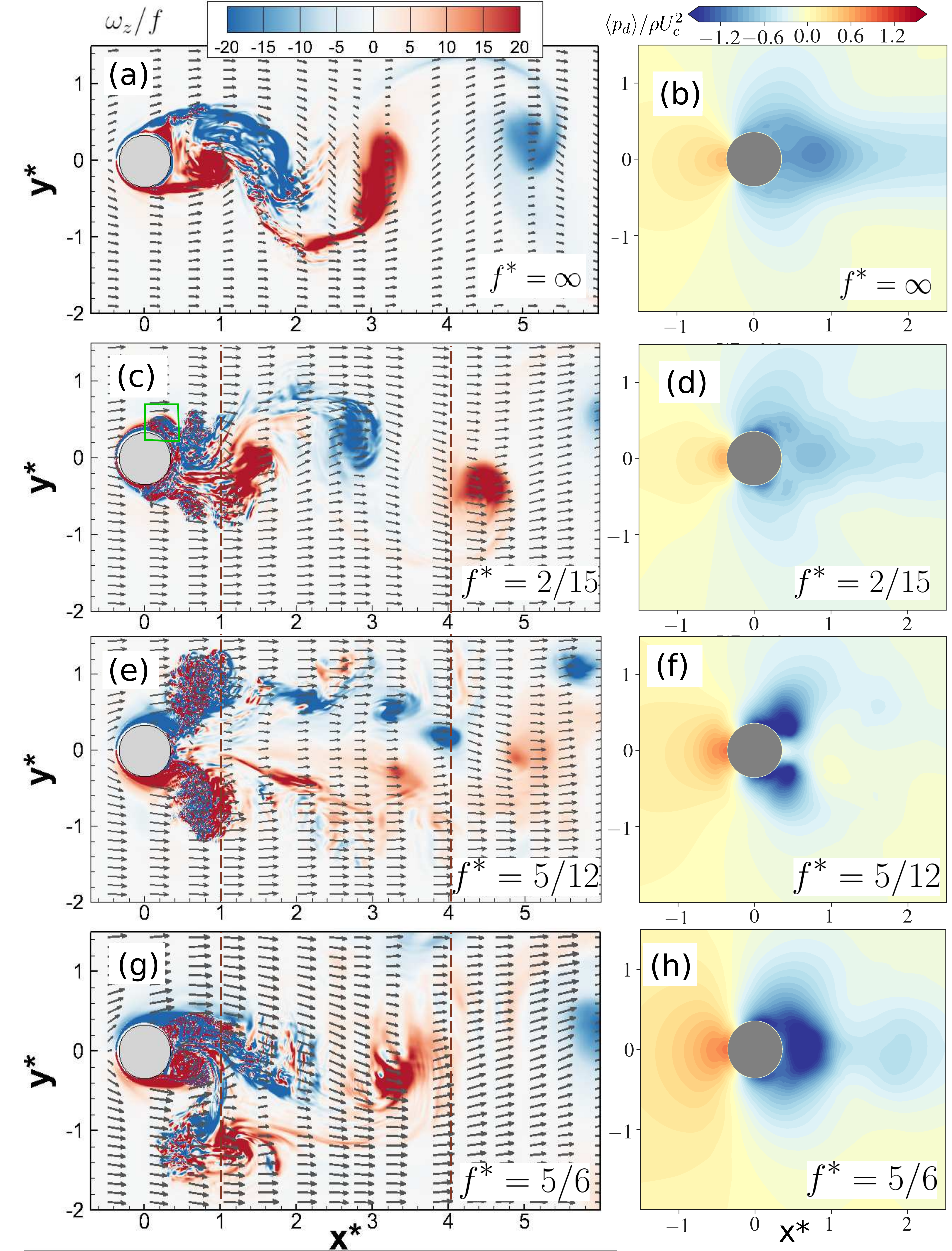}
\caption{
Instantaneous contours of normalized vertical vorticity ($\omega_z/f$) and time-averaged pressure ($\langle p_d \rangle/\rho_0 U_c^2$) in the horizontal plane $z^*=0.25$ : (a,b) $f^*=\infty$, (c,d)   $f^*=2/15$, (e,f) $f^*=5/12$ and (g,h) $f^*=5/6$. In the tidally-modulated cases, the contours of vorticity are plotted at the maximum velocity phase ($\Omega_t t=\pi/2$). 
}
\label{fig:omgx_fstar}
\end{figure}

\begin{figure}[ht!]
\centering
\includegraphics[trim={0.0cm 0.0cm 0cm 0cm},clip, width=0.999\linewidth, angle=0]{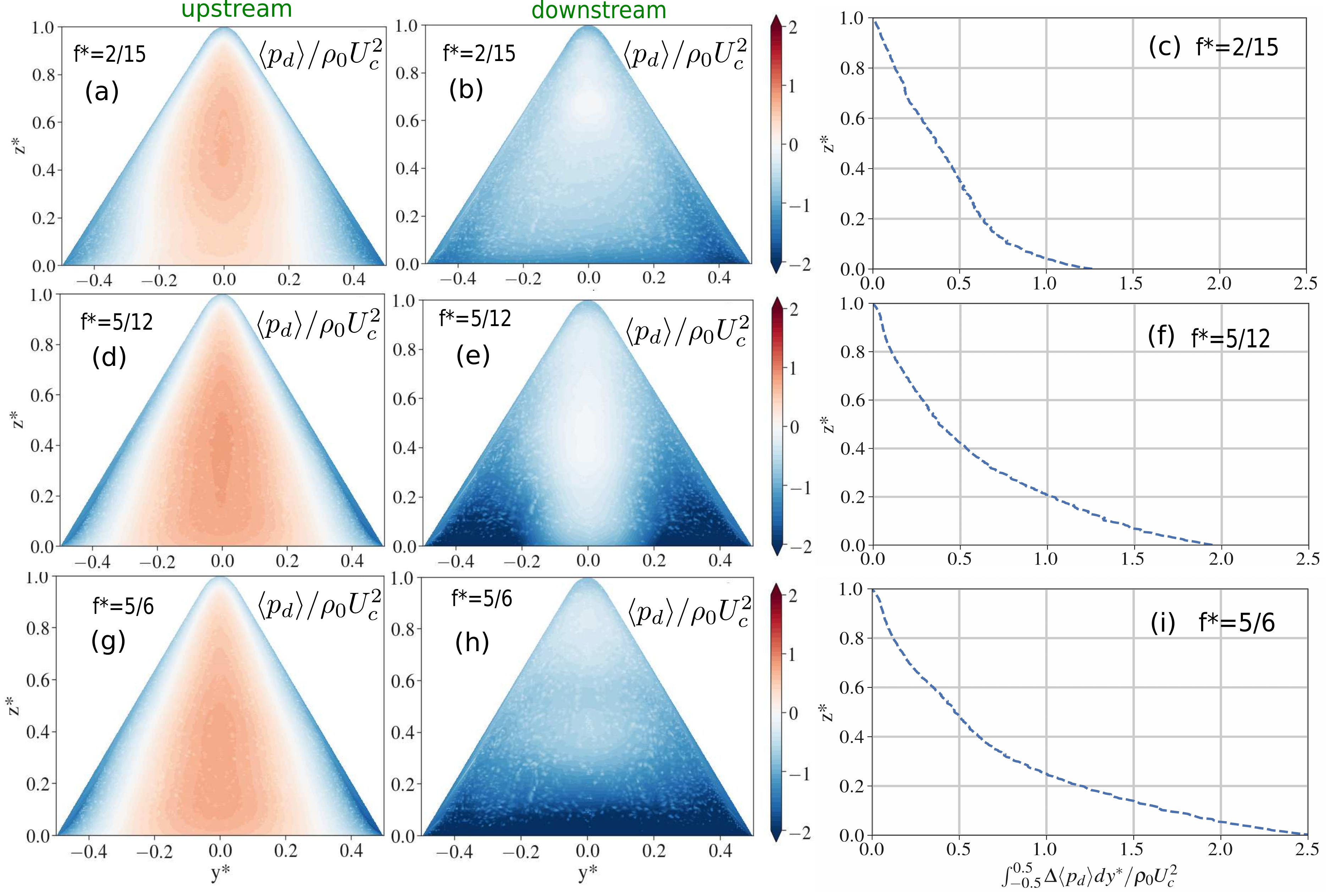}
\caption{
Distribution of mean pressure ($\langle p_d\rangle /\rho_0 U_c^2$) on the upstream and downstream faces of the topography 
: (a,b) $f^*=2/15$, (d,e) $f^*=5/12$ and (g,h) $f^*=5/6$. The difference between upstream and downstream normalized pressure ($\Delta \langle p_d\rangle/\rho_0 U_c^2$) is integrated along $y^*$ and plotted in the last column for each case: (c) $f^*=2/15$, (f) $f^*=5/12$ and (i) $f^*=5/6$.
}
\label{fig:pdiff_body}
\end{figure}

\begin{figure}[ht!]
\centering
\includegraphics[trim={0.1cm 0.0cm 0cm 0.0cm},clip, width=0.9\linewidth, angle=0]{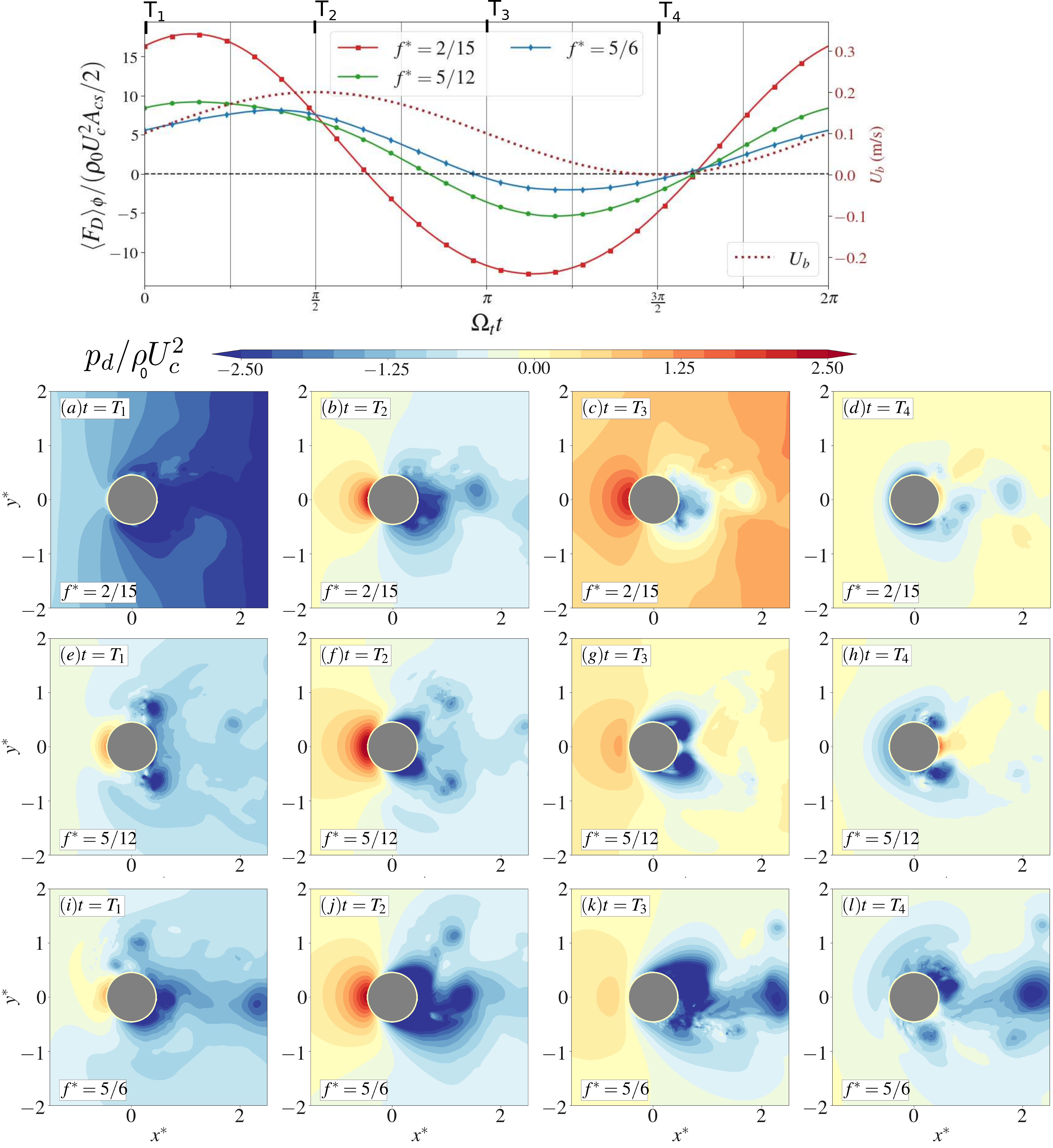}
\caption{
Header: Variation of normalized phase-averaged drag force ($\langle F \rangle_\phi$) with tidal phase $\Omega_t t$, plotted for $f^*=2/15,5/12$ and 5/6 cases. The barotropic velocity $U_b$ is represented by the dotted line \C{in the header}.
The instantaneous contours of dynamic pressure $p_d$ are shown at four phases $T_1, T_2, T_3$ and $T_4$ (marked on the header) for cases $f^*=2/15 (a-d) , f^*=5/12 (e-h)$ and $f^*=5/6 (i-l)$ at $z^*=0.02$. 
The projection of obstacle area in the streamwise direction is denoted as $A_{cs}$.
}
\label{fig:drag_phase}
\end{figure}

\begin{figure}[ht!]
\centering
\includegraphics[trim={0.1cm 0.0cm 0cm 0cm},clip, width=0.999\linewidth, angle=0]{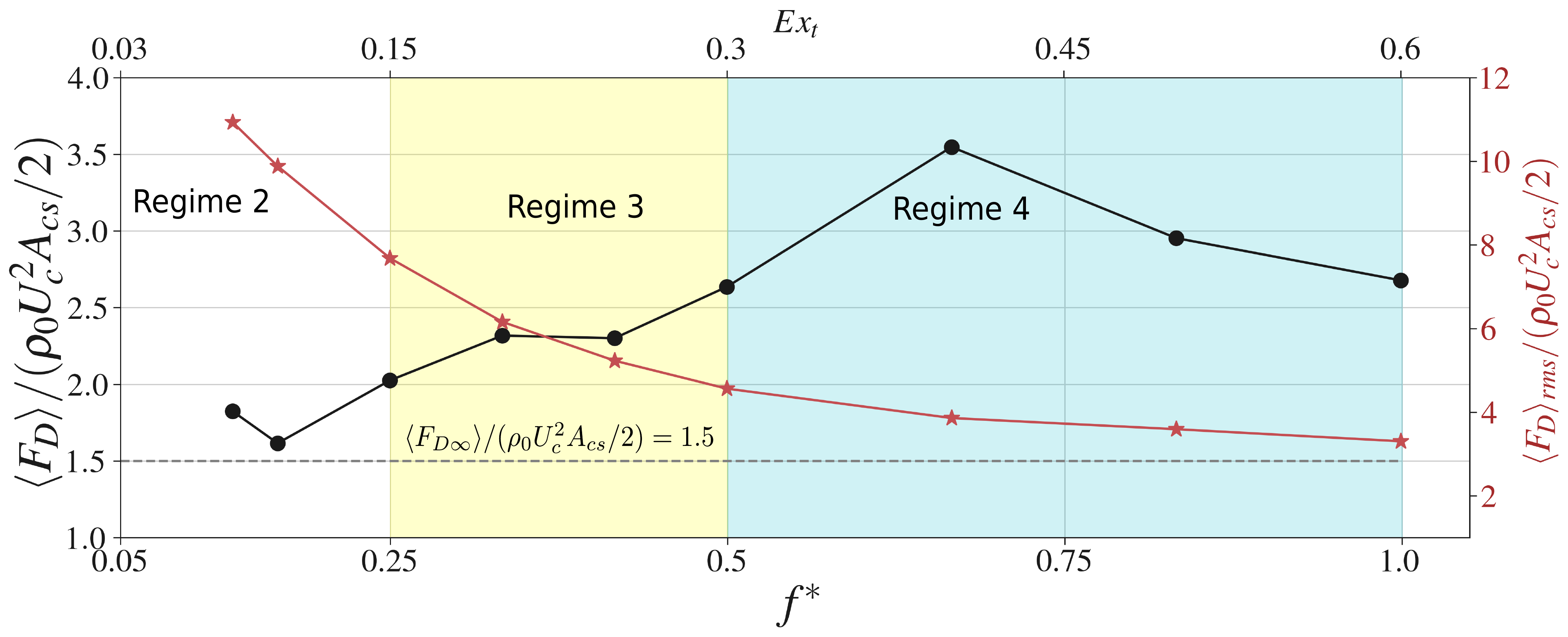}
\caption{
Variation of mean drag force ($\langle F_D \rangle$) and its root-mean-squared value ($\langle F_D \rangle_{rms}$) with $f^*$. The gray-dashed line denotes the normalised mean drag in the $f^*=\infty$ (no-tide) case (represented as $\langle F_{D\infty} \rangle $). 
}
\label{fig:drag_lineplot}
\end{figure}

 \subsection{Mean pressure distribution } \label{subsubsec:pres_dist}


\Cref{fig:omgx_fstar}(a,c,e,g) show qualitative differences in the vortex shedding patterns amongst the cases $f^*=\infty,2/15, 5/12$ and 5/6. On the right (\cref{fig:omgx_fstar}b,d,f,h) the contours of mean dynamic pressure field are plotted. 
 For $f^*=\infty$, shear layers elongate and separate from each side, rolling up into two attached vortices of opposite sign in the lee. 
With time, the attached eddies start to oscillate and eventually break off alternately to form the vortex street in \cref{fig:omgx_fstar}a.
The eddy 
which remains attached briefly, possesses a low pressure core which decays radially outward from the vortex center (not shown). 
The near-wake vertical vorticity values are as large as $20f$, which is in agreement with 
observations of high $Ro_c$ wakes \citep{ChangJLCM_JPO:2019,MacKinnonA_JGR:2019}.

%

When $f^* =2/15$, \Geno{the lateral elongated shear layers are absent} (\cref{fig:omgx_fstar}c). Instead, at this instant, the anticyclonic (negative \C{$\omega_z$}) vortex remains attached to the obstacle as the cyclonic (positive \C{$\omega_z$}) vortex moves into the wake. 
The anticyclonic vortex grows in size from repeated small pulses of vorticity (green box) created every tidal cycle. These pulses form on the lateral sides of the obstacle at the location of flow separation. 
%
The attached vortex is shed at a slower frequency of $f_{s,c}$ \citep{PuthanGRL:2020}. Since $f_{t}\approx 7.5 f_{s,c}$, the eddy remains attached while its circulation increases as the small pulses coalesce 
 over 7.5 tidal cycles. Beyond $x^*=1$, the staggered arrangement of vortices resembles a \Karman{} vortex street. 
The magnitude of mean dynamic pressure $\langle p_d \rangle$ is similar between the  $f^*=\infty$ and $2/15$ cases (\cref{fig:omgx_fstar}b,d). A key difference with respect to \C{the} $f^*=\infty$ case is the presence of low pressure regions on the lateral sides of the obstacle in $f^*=2/15$ (\cref{fig:omgx_fstar}d), created by these vortex pulses.

 The organization of vertical vorticity ($\omega_z$) and mean pressure ($\langle p_d \rangle $) \C{for} $f^*=5/12$ and 5/6 
 are strikingly different from a \Karman{} vortex street generated by a steady current.
\Pranav{At $f^*=5/12$}, a symmetric arrangement of vortices is present in the near wake, attached to the obstacle (\cref{fig:omgx_fstar}e). 
However this configuration is not stable and gives way to a distorted antisymmetric pattern of vortex fragments farther downstream. 
The timing of symmetric twin vortices is phase locked to the barotropic flow, elucidated further in \cref{subsubsec:symm_shedding}. 

\Cref{fig:omgx_fstar}g provides a snapshot of wake when $f^*=5/6$. \C{The} cyclonic vortex remains attached to the obstacle lee \C{at this instant} and  \C{grows in size during the tidal acceleration phase.} 
 Contours of mean pressure reveal regions of large pressure drop in the lee for the $f^*=5/12$ and $f^*=5/6$ cases (\cref{fig:omgx_fstar}f,h). For $f^*=5/12$, 
 two laterally offset low pressure zones lie symmetrically with respect to $y^*=0$ while for $f^*=5/6$, the low pressure region extends to cover the entire rear of the obstacle ($x^*>0$). 
 To discern the pressure anomalies along the vertical extent of the obstacle, the distribution of mean pressure field $\langle p_d \rangle$ \textit{on} the hill is plotted in \cref{fig:pdiff_body}.\\%

 \Cref{fig:pdiff_body}(a,d,g) and  \cref{fig:pdiff_body}(b,e,h) show the pressure distribution on the upstream and downstream faces of the obstacle, respectively.
\C{Each row in \cref{fig:pdiff_body} corresponds to a different case. The 
panels} of \cref{fig:pdiff_body}(c,f,i) \C{contrast} the difference between the upstream and downstream pressure values ($\Delta \langle p_d \rangle$) integrated along $y^*$ \C{among the three cases.}
The distribution of pressure on the upstream face is similar in all three tidally modulated cases. 
However, the differences in flow separation change the pressure distribution on the downstream face. 
%
The eddy shed from the obstacle has a larger horizontal length scale at the boundary as $f^*$ increases.
This forced eddy is characterized by the presence of a low pressure core. 

In all cases, $\Delta \langle p_d \rangle$ is bottom intensified. 
\C{In their} steady-current simulations,
 \cite{MacCready:2001} noted  ``a tendency for drag on the lower half of the ridge to be greater than that on the upper half". \C{In the present tidally modulated cases}, the bottom intensification progressively increases from $f^*=2/15$ to $5/6$.
%
 \C{For example}, consider the laterally integrated mean pressure at height $z^*=0.2$ (\cref{fig:pdiff_body}c,f,i). Its value increases from 0.6 at $f^*=2/15$ to 1 at $f^*=5/12$.
 For the $f^*=5/6$ case (\cref{fig:pdiff_body}h,i), 
 the near-bottom integrated $\Delta \langle p_d \rangle$ 
is up to five times larger \C{than} its value 
at $z^*=0.5$.
A similar bottom-intensified pressure difference is also observed for $f^*=5/12$, though for this case it is confined to the lateral sides of the abyssal hill (\cref{fig:pdiff_body}e,f). 
%
%
%
Even for $f^*=2/15$ (\cref{fig:pdiff_body}b,c), a {twofold} increase in pressure drop is observed near the bottom relative to the upper half of the obstacle. 

The eddy shedding (\cref{fig:omgx_fstar}) and pressure anomalies (\cref{fig:pdiff_body}) in \C{the} $f^*=2/15$, 5/12 and 5/6 cases are characteristic of the regimes 2, 3 and 4 which they represent. 
 An examination of the pressure variability within a tidal cycle is the next step towards 
   revealing the underlying mechanisms governing enhancement of drag at the obstacle base. 

%
\Cref{fig:drag_phase} shows snapshots of instantaneous normalized pressure $ p_d/\rho_0 U_c^2 $ in the horizontal plane $z^*=0.02$ (close to the bottom boundary) for $f^*=2/15$ 
, 5/12 and 5/6 cases. Four time instants $T_1, T_2, T_3$ and $T_4$  are chosen at tidal phases $\Omega_t t=0,\pi/2,\pi$ and $3\pi/2$, respectively, for the snapshots. 
The variation of the phase-averaged form drag $\langle F_D\rangle_\phi$ and the barotropic current $U_b$ are plotted in the header.

Consider the case $f^*=5/6$ (from regime 4). At $t=T_1$, the accelerating fluid impinges on the obstacle, creating high pressure upstream and a low pressure zone downstream (\cref{fig:drag_phase}i). Owing to the asymmetry in flow separation, the low pressure region is dominant between $y^*=-0.5$ and 0.
As the barotropic velocity increases, the region of low pressure enlarges downstream in \cref{fig:drag_phase}j, associated with the lee eddy formation illustrated in \cref{fig:omgx_fstar}g. The simultaneous increase in upstream pressure heightens the pressure anomaly between the fore and lee. 
As the barotropic flow decelerates between $T_2$ and $T_3$, the upstream pressure recedes concomitantly while the low pressure
 in the lee is sustained at $t=T_3$ (\cref{fig:drag_phase}k). The lee eddy remains attached at this instant preserving the low pressure zone.
The low pressure region continues to exist at the downstream face between $y^*=0$ to 0.5 at $T_4$ (\cref{fig:drag_phase}l).
With the formation of persistent anticyclonic and cyclonic vortices on opposite sides during successive tidal cycles, the low pressure in the lee is maintained, albeit subject to lateral oscillations. 
The outcome is an \Geno{elevated} mean drag. 
 %



For case $f^*=5/12$, the pressure drop downstream of the obstacle is symmetric  owing to the nature of the vortex shedding for this case. The low pressure region begins to form immediately after the lateral separation of the eddies at $t=T_1$ on both 
 sides of the obstacle lee (\cref{fig:drag_phase}f). While the barotropic current remains above its mean value $U_c$ between $t=T_1$ and $T_3$, the vortices continue to grow larger on either side of the hill (elaborated further in \cref{subsec:vortex_dynamics} and \cref{fig:symm_shedding}).
The pressure drop across the hill increases concurrently in \cref{fig:drag_phase}f and \cref{fig:drag_phase}g. 
At the zero velocity phase ($t=T_4$), the recirculating fluid at low pressure accelerates upstream relative to the barotropic flow. 
A region of positive dynamic pressure (\cref{fig:drag_phase}h) is identified at this instant near $x^*=0.5 $ and $y^*=0$, while the pressure upstream at the obstacle centerplane ($x^*=-0.5 $ and $y^*=0$) drops below \Geno{zero} momentarily. The eddy-induced low pressure zone remains attached at $x^*,y^*=(0.5,\pm 0.5)$. 
This configuration is reset to \cref{fig:drag_phase}e as the background flow gains momentum in the next cycle.
%
%

The instantaneous dynamic pressure $p_d$ has contributions from two sources: the added mass and flow separation (see \cref{eq:drag3}). When the tidal frequency is larger, the added mass component increases owing to its dependence on the tidal acceleration magnitude $U_t\Omega_t$.
 This is likely to occur when $f^*=2/15$ \C{as corroborated by}
 normalised $p_d$ in \cref{fig:drag_phase}a-d. 
%
  At $T_1$, the tidal acceleration reaches its maximum value and the dynamic pressure $p_d/\rho_0 U_c^2$ drops below $-2.5$ over a large region in the obstacle lee (\cref{fig:drag_phase}a).
   Similarly, at the phase of maximum tidal deceleration ($t=T_3$), $p_d/\rho_0 U_c^2$ exceeds $2.5$ over a large region upstream of the obstacle (\cref{fig:drag_phase}c). At instances of zero acceleration ($t=T_2$ and $T_4$), the dynamic pressure is purely associated with flow separation. At $T_2$, the low pressure region in the lee (\cref{fig:drag_phase}b) is generated due to the attached eddy in \cref{fig:omgx_fstar}c, while the high $p_d$ upstream is generated by the impinging current. 
   On the contrary, at $T_4$, the pressure anomalies in the obstacle centerplane are marginal (\cref{fig:drag_phase}d). \C{At the lateral sides of the obstacle}, a weak pressure drop is observed, attributed to the vortex pulses discussed earlier.
  The tidal period is smaller than the natural eddy shedding time scale by a factor of 7.5 for this case. 
  This constrains the size of the attached eddy to a small vortex pulse, \C{which has a  \Geno{correspondingly} weak effect on the pressure drop. 
  }
  
 The \C{temporal} variability of $p_d$ directly affects the \Geno{variation in} form drag \C{as illustrated by} the phase-averaged form drag $\langle F_D\rangle_\phi$ in the header of \cref{fig:drag_phase}.
 %
  The form drag exhibits large modulation in the $f^*=2/15$ case. Also, the phase difference between $\langle F_D\rangle_\phi$ and $U_b$ exceeds $\pi/4$. 
  Recall that the inertial drag has a phase difference of $\pi/2$ with $U_b$. Therefore, it is possible that instantaneous inertial drag contributions are important in this case. 
  To confirm this, we follow the procedure of \cite{WarnerM_JPO:2009} to estimate the ratio of inertial drag to separation drag for our obstacle geometry.
  We assume \Geno{that} the magnitude of $F_D^S$ is equal to the bluff-body drag estimate of $\rho_0 A_{cs} U_c^2/2$ and \Geno{that} $C_a=1$ in \cref{eq:added_mass} (for explanation, see pg. 2979 of \cite{WarnerM_JPO:2009}). 
  For the conical obstacle, and taking $R=U_t/U_c=1$,
  \begin{align}
 \frac{| F^I_D| }{|F_D^S|} = \frac{2\pi}{3Ex_t} \approx \frac{3.5}{f^*}  ,
  \label{d:ratio}
  \end{align}
  %
  where the second equality follows from $f^*=1.66Ex_t$ from \cref{intro}.
   As $f^*$ increases, the inertial component of form drag decreases. Separation drag becomes more prominent relative to inertial drag. This makes sense because at larger $f^*=f_{s,c}/f_t$, 
 \Geno{the tidal acceleration is weaker which causes the magnitude of inertial drag to decrease.} 
   Thus the phase difference between $\langle F_D\rangle_\phi$ and $U_b$ is \Geno{smaller than $\pi/2$} in $f^*=5/12$ and 5/6.

 \subsection{States of high drag} \label{subsubsec:drag_amp}
  From the preceding text, it is clear that the pressure anomalies on the obstacle change significantly owing to tidal oscillations. Here we demonstrate that tidal oscillations lead to high levels of instantaneous and mean drag.
 The ratio of inertial drag to separation drag is inversely related to $f^*$ as shown in \cref{d:ratio}. It varies from a value of approximately 4 at $f^*=5/6$ to 25 at $f^*=2/15$. Clearly the contribution of inertial drag exceeds the separation drag within the tidal cycle, especially for low $f^*$ cases.
%
\Cref{fig:drag_phase} showed that normalised $\langle F_D \rangle_\phi$ in $f^*=2/15$ reached values of up to 16 during the tidal cycle just after $t=T_1$ and dropped rapidly below -10 near $T_3$. 
%
On the other hand, $\langle F_D \rangle_\phi$ has a smaller variance over the tidal cycle in $f^*=5/12$ and 5/6.
\C{In these two cases, large eddy cores in the lee generate zones of low pressure with high magnitude which 
 remain at the obstacle periphery throughout the tidal cycle (\cref{fig:drag_phase}e-l).} The longer residence time of the attached low pressure zone enhances 
the separation drag $F_D^S$. 
The deviation of $\langle F_D \rangle_\phi$ from its mean value is also much smaller. The values of $\langle F_D \rangle_\phi$ lie between -5 and 8 for this case. 

Quantifying the mean and RMS values of $F_D$ is key in this comparative analysis.
The time averaged value, namely $\langle F_D\rangle$, offers a diagnosis of high-drag states by nearly eliminating the contribution from inertial drag 
 and providing an estimate of the separation drag $F_D^S$.
 Parametrization of separation drag is important owing to its ability to extract energy from the flow. 
%
 To this end, the normalized $\langle F_D\rangle$ and $\langle F_D\rangle_{rms}$ are plotted in \cref{fig:drag_lineplot}
  \C{as a function of} $f^*$. \Sutanu{The normalized  form drag or the drag coefficient  is  $\langle F_D\rangle/  (\rho_0 U_c^2 A_{cs}/2)$ where $A_{cs}$ is the 
  projected  obstacle area in the streamwise direction.}  
The value of form drag coefficient in \C{the} $f^*=\infty $ case is $1.5$. 
In the tidally modulated cases, the mean form drag \Sutanu{coefficient} \C{generally increases} with increasing $f^*$. The larger size and longer residence time of the attached wake eddy contribute to the  intensification of \C{the} pressure drop and its sustenance near the foot of the abyssal hill (\cref{fig:pdiff_body}h,i). 
%
%
 As a result, a large increase in $\langle F_D\rangle$ (relative to $\langle F_{D\infty} \rangle $).
 is observed in regimes 3 and 4. The mean drag coefficient exceeds 3.5 in regime 4, and \C{its} average value is $ 3$, signifying a \C{two-fold increase in form drag with respect to the steady case in regime 1.}
 In regime 3, 
 there is a 60\% increase in \C{average} form drag \C{relative} to $\langle F_{D\infty}\rangle$.
 In comparison, only a marginal increase of $\langle F_{D}\rangle$ is seen in regime 2. 
 On the other hand, the RMS drag, $\langle F_{D}\rangle_{rms}$ is \Geno{strongly} influenced by inertial forces. 
  The RMS drag decays rapidly as $f^*$ increases, from a large value of 11 at $f^*=2/15$ to 3.2 at $f^*=1$. This monotonic decay is attributed to a reduction in the inertial drag associated with the tidal acceleration.
 %
 
 To ascertain the magnitude of form drag relative to the drag associated with the frictional bottom boundary layer (BBL), consider \C{regime 4}. The mean form drag \C{$\langle F_{D\infty}\rangle=3\rho_0 U_c^2 A_{cs}/2$}. Assuming a coefficient of bottom friction ($C_f$) of approximately $0.002$ based on previous studies \citep[e.g.][]{McCabePP_JPO:2006}, the mean frictional drag ($\langle F^{BBL} \rangle$) is of the order $C_f\rho_0 U_c^2 A_b/2$, where $A_b$ is the projected area of the hill on the horizontal plane. Therefore, the ratio of mean form drag to frictional drag is 
 \begin{align}
 \frac{\langle F_{D\infty}\rangle}{\langle F^{BBL} \rangle} = \frac{3 A_{cs}}{C_f A_b} \approx 300 
 \label{eq:form_friction} 
\end{align}  
 %
 \C{ Therefore, it is important to account for the form drag of 3D underwater obstacles when parametrising bottom drag, especially when tidal oscillations are present.
 }
The interplay between form drag and spatio-temporal organisation of the lee eddies deserves attention. To this end,  vortex dynamics are explored in the next section.
 

%% file: vortex_dynamics.tex
Investigations of 
cylinder wakes created by homogeneous non-rotating flow, have often been used to add to our understanding of vortex dynamics in oceanic wakes \citep{ChangJLCM_JPO:2019}. 
 However, density stratification and planetary rotation influence the wake significantly. For example, \cite{LinP_ARFM:1979} presented a detailed review of internal wave radiation and \C{the} emergence of thin lee vortices when $Fr_c<1$. 
With \C{the} addition of rotation, \cite{DongMS_JPO:2006} showed manifestations of barotropic, baroclinic and centrifugal instabilities in the wake.
 The recent study of \cite{PuthanGRL:2020} showed that vortices \C{in the far wake} occur at frequencies coinciding with tidal subharmonics. 
 \C{Form drag is directly related to flow separation and near wake vortices. We discuss these facets and also the vertical organization of lee vortices in this section. }
\subsection{Vertical structure of lee vortices} \label{subsubsec:Lee_wake_vortices}

\begin{figure}[ht!]
\centering
\includegraphics[trim={0cm 0cm 0cm 0cm},clip, width=1.1\linewidth, angle=0]{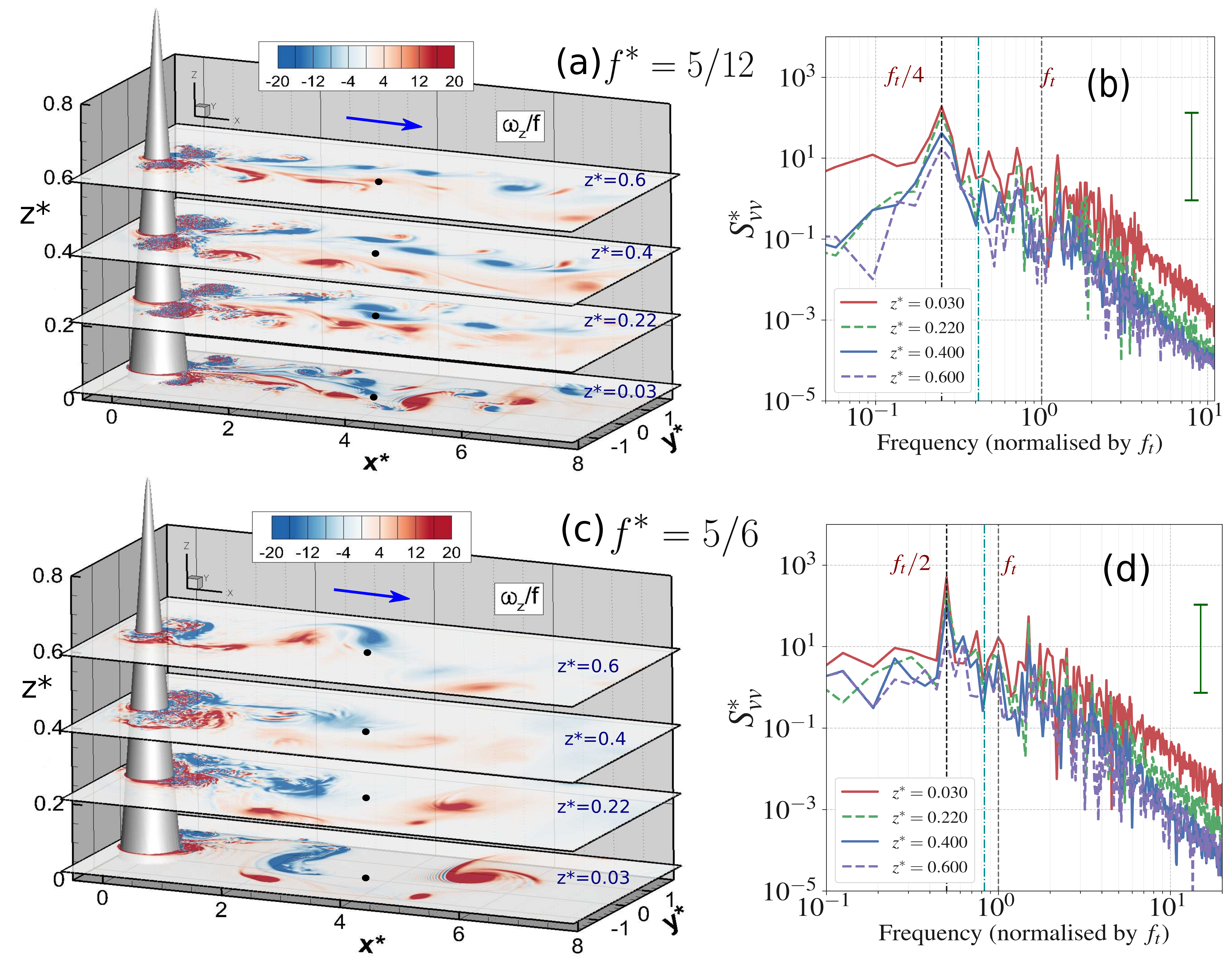}
\caption{
Eddy formation is depicted by the normalized vertical vorticity ($\omega_z/f $) in four horizontal (x-y) planes : (a) $f^*=5/12$ and  (c)$f^*=5/6$. (b,d) Spectra of spanwise velocity $S_{vv}^*=S_{vv}/U_c^2$, plotted at four probes (black dots in a,c respectively) 
 chosen at different vertical heights located at $x^*=4$ and $y^*=0$. The green bar in (b,d) shows the 95\%
confidence interval of the spectrum.
}
\label{fig:omgx_vertical_planes}
\end{figure}
\Cref{fig:omgx_vertical_planes}a,c show contours of $\omega_z$ on horizontal planes at four different heights in the wake for $f^*=5/12$ (regime 3) and 5/6 (regime 4). Velocity spectra at spatial probes chosen at these vertical heights are shown in \cref{fig:omgx_vertical_planes}b,d. Let $z^{*}_{div} \approx 1-(U_c+U_t)/Nh = 0.7$ be \C{an} approximate estimate of the dividing vertical height, below which the fluid is driven laterally around the obstacle \citep{Sheppard_QJRM:1956,Drazin_T:1961}. The four vertical heights in \cref{fig:omgx_vertical_planes} are chosen such that $z^*<z^*_{div}$.

For $f^*=5/12$ (\cref{fig:omgx_vertical_planes}a),  laterally symmetric dipoles, attached to the obstacle, form along its vertical extent. 
 Farther in the wake beyond $x^*=2$, discrete eddies remain scattered and disorganized. \Geno{At this stage,} these eddies do not exhibit the spatial configuration 
 of a \Karman{} vortex street. 


For the case of $f^*=5/6$ (\cref{fig:omgx_vertical_planes}c), the coherent vortices are larger and their shedding is asymmetrical.  A region of both negative and positive vorticity remains attached to the topography along its vertical extent. Beyond $x^*=1$, a large cyclonic (positive) vortex is followed by a large anticyclonic (negative) vortex at $z^*=0.03$ and 0.22. These two vortices are found to be aligned vertically up to a height of $\mathcal{O}(U_t/N)$. 
Similar aligned vortices were also observed in the $f^*=5/12$ case, albeit smaller in diameter and more scattered. 
 \C{A} possible explanation may stem from the vortex separation \C{being} initiated by the tidal forcing. 
To illustrate this, the spectra of spanwise velocity ($S_{vv}^*$) \C{are} plotted at different vertical heights for $f^*=5/12$  and $f^*=5/6$ cases in \cref{fig:omgx_vertical_planes}b,d. The frequency of lateral oscillations from  $S_{vv}^*$ is used to ascertain the vortex shedding frequency.
%
Vortices are shed at the same instant during the tidal cycle along the vertical extent of the obstacle (not shown). In addition, in the far wake at $z^* < z^*_{div}$, the vortices are observed 
at a uniform frequency of $f_t/4$ in $f^*=5/12$ case and $f_t/2$ in $f^*=5/6$ case. 
%
 The timing of the flow separation may help align the vortices along the $z$ direction up to a vertical length scale permitted by the background stratification. 
 At elevations above $z^*_{div}$, the transient lee waves interfere strongly with the coherent structures in the wake \citep[see][]{PuthanOM:2020}. 
Near wake turbulence creates small-scale patches of vorticity in both cases, up to two diameters away from the obstacle. However, we emphasize that their instantaneous presentation is qualitatively different from a classical \Karman{} vortex street over the vertical extent of the hill.
%


%% file: symmetric_vortices.tex
\begin{figure}[ht!]
\centering
\includegraphics[trim={0cm 0cm 0cm 0cm},clip, width=0.999\linewidth, angle=0]{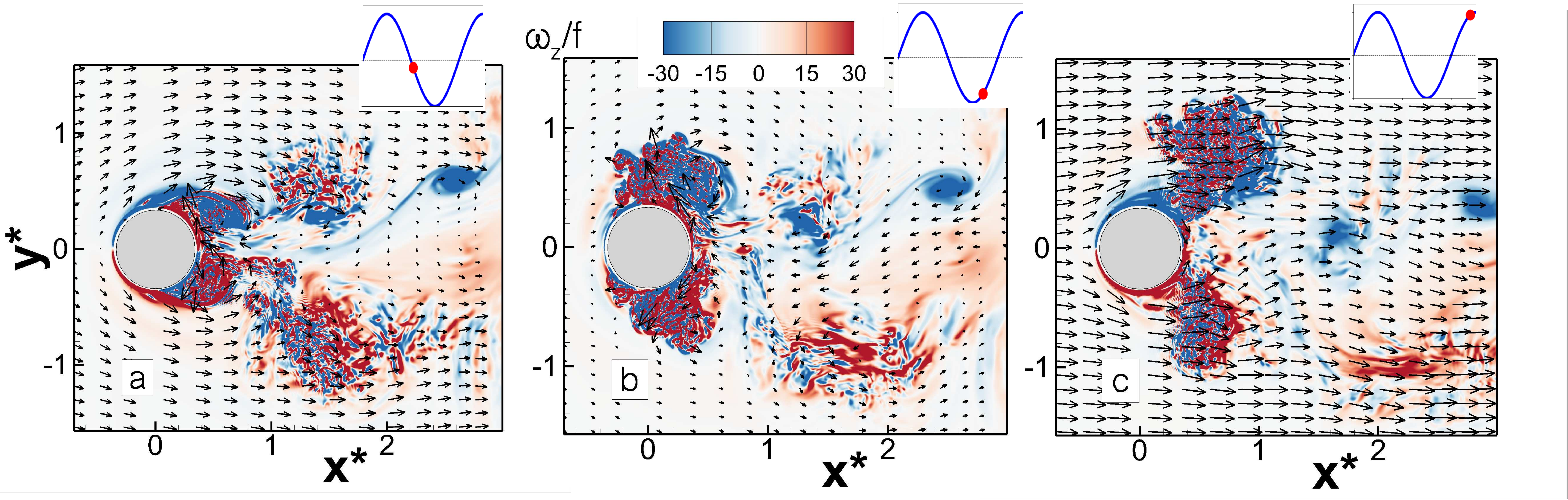}
\caption{
Normalized vertical vorticity ($\omega_z/f$) at three different phases of a tidally perturbed wake at $f^*=5/12$ : (a) $t/T = 18.52$, (b)$t/T= 18.85$ and (c) $t/T=19.19$.
}
\label{fig:symm_shedding}
\end{figure}

The symmetric twin dipoles \C{shown previously} in \cref{fig:omgx_fstar}e are created through a sequence of events presented in \cref{fig:symm_shedding}.
A locally adverse streamwise pressure gradient (shown in \cref{fig:drag_phase}e,f) develops on either lateral side of the obstacle. While the barotropic flow remains positive, as in \cref{fig:symm_shedding}a, two opposite-signed vortices form in the recirculation zone and grow in size until the velocity approaches zero. 
As the tide-associated pressure gradient changes sign at the zero velocity phase, the high vorticity fluid accelerates upstream relative to the background flow on both lateral sides. During this event, additional vorticity of opposite sign is generated from shear when this fluid is near the obstacle. 
For example, the attached anticyclonic eddy in \cref{fig:symm_shedding}a accrues positive vorticity in \cref{fig:symm_shedding}b during its deflection to the $+y$ direction. 
Thus twin dipoles form symmetrically on either lateral side of the obstacle.
%
During the subsequent acceleration phase, the dipoles gain enough momentum to advect downstream (\cref{fig:symm_shedding}c). At the same instant, a new vortex pair starts to grow near the separation points, completing a full cycle of oscillation.

The formation of the symmetric vortices are phase locked to the tidal cycle. 
Note that the positive vorticity generated by the upstream flow (directed to the $+y$ direction) in the previous tidal cycle (located at $x^*,y^*=1.5,0.5 $) in \cref{fig:symm_shedding}a, decays in the present tidal cycle, as shown in \cref{fig:symm_shedding}b,c. 
%
%
The vortices also lose their lateral symmetry beyond $x^*=1.5$. 
Tidal forcing aids in the symmetrical placement of vortices up to an excursion distance of $U_t/f_t$, or equivalently, $x^*=1.57$ for this case.
 Beyond $x^*=1.57$, a staggered and distorted pattern of vortices occurs (see \cref{fig:omgx_fstar}e). Deviation of the wake vortex frequency $f_{s}$ from the tidal frequency $f_t$ occurs in this region, as elaborated in the next section.

%% file: Temporal_vort.tex
\begin{figure}[ht!]
\centering
\includegraphics[trim={0.1cm 0.1cm 0.1cm 0cm},clip, width=0.999\linewidth, angle=0]{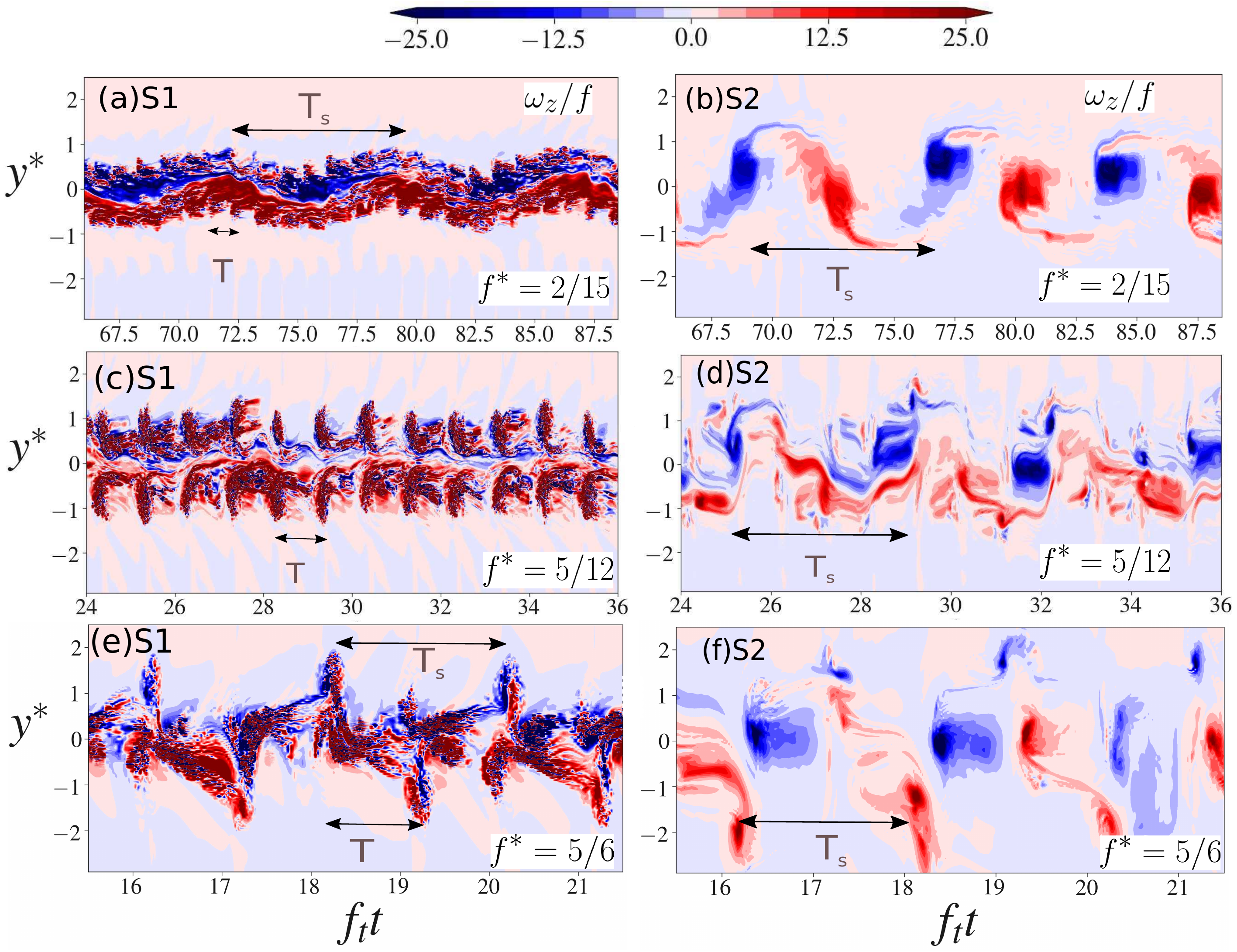}
\caption{
Time evolution of vertical vorticity ($\omega_z/f$) is depicted by a $y-t$ Hovm\"{o}ller diagram, at stations S1 (at $x^*,z^*=1,0.25$) and S2 (at $x^*,z^*=4,0.25$) for three cases : (a,b) $f^*=2/15$,  (c,d) $f^*=5/12$ and (e,f) $f^*=5/6$. Here, $T$ is the tidal period and $T_s$ is the time period \C{of far wake vortices}. Note that the range of $f_tt $ decreases from the top to the bottom row. The stations S1 and S2 are indicated by brown dashed lines in \cref{fig:omgx_fstar}.
}
\label{fig:hof_plots}
\end{figure}


\begin{figure}[ht!]
\centering
\includegraphics[trim={0.1cm 0.1cm 0cm 0cm},clip, width=0.999\linewidth, angle=0]{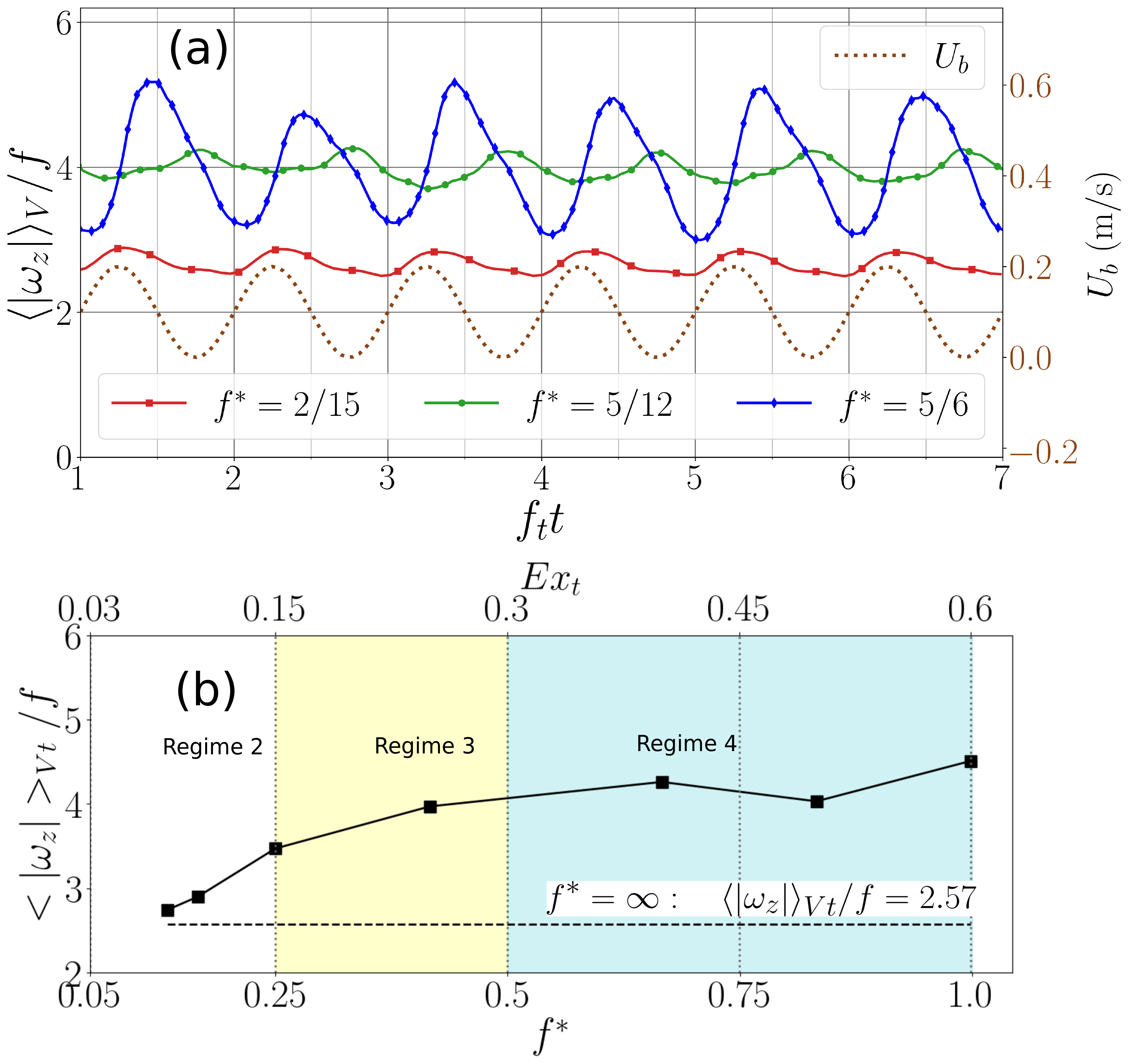}
\caption{
(a) Volume-averaged vertical vorticity $\langle | \omega_z |\rangle_{V}=\int_0^V |\omega_z|dV/V$ plotted as a function of time for $f^*=2/15,5/12$ and 5/6 cases. (b) Variation of time averaged $\langle |\omega_z| \rangle_{V}$ shown \C{as a function of $f^*$}. 
}
\label{fig:vorticity_integ}
\end{figure}


Hovm\"{o}ller diagrams of normalized vorticity in \cref{fig:hof_plots} demonstrate \C{clearly that transitions of the vortex frequency \Geno{vary} from the near to the far wake between regimes 2 to 4. }
%
 Space-time ($y-t$) contours of vertical vorticity are plotted at two stations: S1 in the near wake and S2 in the far wake \Geno{(see dashed lines in \cref{fig:omgx_fstar}c,e,g)}. 
In \cref{fig:hof_plots}a, the signature of the tidal frequency is observed in the vortex pulses spaced over one tidal period ($T$) for $f^*=2/15$. These pulses are arranged in the form of a slowly varying large-scale 
 sinuous (antisymmetric) mode.
The sinuous mode evolves into a row of opposite-signed vortices at S2 (\cref{fig:hof_plots}b). 
 The signature of the individual vortex pulses disappears \C{at S2} leaving large-scale coherent vortices separated by the time period, $T_s$. 
 The vortex period ($T_s$) in the far wake aligns with the natural shedding period of the obstacle wake, $T_{s,c}$. 

Case $f^*=5/12 $ exhibits a laterally symmetric vortex pattern at S1, depicted in  \cref{fig:hof_plots}c. These vortex structures repeat every tidal cycle.  
%
 The unstable symmetric mode transitions into an antisymmetric downstream wake made up of coherent vortices in \cref{fig:hof_plots}d.
 These vortices are separated by $T_s=4T$, demonstrating the tidal synchronization \C{noted} in \cite{PuthanGRL:2020}. 
 Thus the wake vortex frequency is modified to $f_t/4$. In other words, during every tidal cycle the symmetric vortices feed their vorticity into a larger vortex downstream, formed every four tidal cycles.

In the $f^*=5/6$ case, the wake is strongly asymmetric in the lateral direction at S1 (\cref{fig:hof_plots}e). Large vortex lobes repeat once every two cycles and extend laterally. 
 The lateral extent stretches farther at S2 relative to the previous two cases, as seen in \cref{fig:hof_plots}f. The wake width is as large as 4.5D (extending from $y^*\approx -2.5$ to $y^* \approx 2$), compared to $3D$ and $2.2D$ for $f^*=5/12$ and $2/15$.
A temporally distorted wake is observed at S2 in the $f^*=5/6$ case, wherein the larger patches of negative vorticity are observed near the center while smaller vortices of same sign are laterally offset by approximately $2D$ from the centerplane. 
%
The space-time plot reveals thin filaments of positive vorticity in the temporal frame, in the far wake of $f^*=5/12$ and 5/6 cases.
The far wake eddy frequency coincides with $f_t/2$ for $f^*=5/6$. 
To explore the time evolution of eddy vorticity injected into the wake, the absolute value of $\omega_z$ is volume averaged over a domain encompassing the hill and extending to a distance of $8D$ into the wake.
\Cref{fig:vorticity_integ}a shows the temporal evolution of normalized volume-averaged vorticity $\langle |\omega_z|\rangle_V/f $. 
The dotted line shows the variation in the barotropic velocity.
For \C{the} $f^*=2/15$ case, $\langle |\omega_z|\rangle_V $ \C{varies} in phase with $U_b$. The pulses of vorticity which form during the maximum velocity phase of every tidal cycle are likely responsible for the small increases of $\langle |\omega_z|\rangle_V $ from its mean value.
At $f^*=5/12$, $\langle |\omega_z|\rangle_V $ increases at the low-velocity phase during every tidal cycle. During the shedding of symmetric vortices, the low velocity phase is accompanied by additional vorticity from shear layer roll-up at the obstacle (\cref{fig:symm_shedding}). The result is an increase of $\langle |\omega_z|\rangle_V $ near the zero-velocity phase.
On the other hand, larger variations from the mean occur in $\langle |\omega_z|\rangle_V $  values when $f^*=5/6$. 

Time averages \C{of $\langle |\omega_z|\rangle_{V}/f$} are plotted over a range of $f^*$ values in \cref{fig:vorticity_integ}b. 
\C{Net} vorticity added by the tides is not significant when $f^*$ lies below $ 0.2$ (\cref{fig:vorticity_integ}b). Above this threshold, a gradual increase is noticed in $\langle |\omega_z|\rangle_{Vt} $ until it reaches a value of 4 at $f^*=5/12$ and does not increase appreciably beyond that.
This may be explained as follows.
Over a tidal cycle, high vorticity fluid in the obstacle vicinity is advected by a distance of $U_c T$. This fluid migrates into the wake permanently if $U_c T$ exceeds $\mathcal{O}(D)$. In other words, the tidal flow adds vorticity to the wake when $U_c T/D > \mathcal{O}(1)$. Simplifying using the relation $f^*=1.66Ex_t$ (from \cref{intro}) this condition is equivalent to $f^*> \mathcal{O}(0.25)$. 
At larger $f^*=f_{s,c}/f_t$, the vortices have sufficient time to separate from the obstacle before the deceleration of barotropic current. 

%
